\documentclass[aps,prl,twocolumn,longbibliography,footinbib,groupedaddress,
]{revtex4-2}

\usepackage{graphicx}% Include figure files
\usepackage{subcaption}
\usepackage{dcolumn}% Align table columns on decimal point
\usepackage{bm}% bold math
\usepackage{amsmath}
\usepackage{amssymb}
\usepackage{color}
\usepackage{hyperref}% add hypertext capabilities
\usepackage{ragged2e}
\usepackage[normalem]{ulem}

\begin{document}

\preprint{APS/123-QED}

\title{Dimensional Control of the Coherence Time of Scattered Light in Cold Atom Clouds}% 

\author{Ana Cipris$^{1,3}$}
\author{Mateus A. F. Biscassi$^{1,2}$}
\author{J. C. C. Capella$^{1,4}$}
\author{Martial Morisse$^{1}$}
\author{Hani Naim$^{1}$}
\author{Hugo Sedlacek$^{1}$}
\author{Apoorav Singh Deo$^{1}$}
\author{Stephan Asselie$^{1}$}
\author{Robin Kaiser$^1$}
\author{Raul Celistrino Teixeira$^2$}
\author{Romain Bachelard$^2$}
\author{Mathilde Hugbart$^1$}  \email{Contact author: mathilde.hugbart@univ-cotedazur.fr}

\affiliation{$^{1}$Universit\'e C\^ote d'Azur, CNRS, INPHYNI, France}
\affiliation{$^{2}$Departamento de F\'{\i}sica, Universidade Federal de S\~{a}o Carlos, Rodovia Washington Lu\'{\i}s, km 235 - SP-310, 13565-905 S\~{a}o Carlos, SP, Brazil}
\affiliation{$^{3}$Instituto de Física de São Carlos, Universidade de São Paulo, São Carlos SP 13566-970, Brazil}
\affiliation{$^{4}$Departamento de F\'{\i}sica, Universidade Federal de Pernambuco, 50670-901 Recife, Pernambuco, Brazil}
\date{\today}% It is always \today, today,

\begin{abstract}
%Cold atomic clouds are promising platforms for generating correlated photons, although broadening mechanisms compromise their temporal coherence. In this work, we demonstrate that the geometry of the cloud can serve as a control parameter for the coherence time of scattered light. Using polarization optics, we monitor the balance between single and multiple scattering as the cloud geometry is tuned from 3D to quasi-1D, while intensity correlation measurements reveal the increase in coherence time. Our results provide a new tool for controlling the temporal coherence of light, with potential applications in quantum optics and communication.
Cold atomic clouds are promising platforms for generating correlated photons, but multiple scattering and associated Doppler broadening limit their temporal coherence. Here we demonstrate that cloud geometry provides a powerful means to extend the coherence time of scattered light. In the experiment, intensity-correlation measurements show that an elongated (quasi-1D) cloud exhibits systematically longer coherence times than a spherical (3D) cloud of the same on-axis optical thickness, as a direct consequence of the suppression of multiple scattering in the elongated geometry. Random-walk simulations reproduce this trend and further show that elongation drives the coherence time toward the single-scattering limit. The combined results establish cloud geometry as a robust control parameter for temporal coherence in cold-atom ensembles, with potential applications in quantum optics and communication.
\end{abstract}

\maketitle

%\tableofcontents

\textit{Introduction.} The ability to control the coherence properties of light is essential for the development of quantum technologies, such as quantum communication, quantum computing, and high precision metrology~\cite{Unruh1995,DiVincenzo1999,Halder2008,Wang2021}. In many applications, the coherence time of light directly affects the efficiency and reliability of quantum state transfer, the precision of measurements, and the scalability of quantum networks~\cite{Lvovsky2009,Norcia_2016}. Consequently, controlling and extending the coherence time of light is a central goal for quantum technologies.

Among the quantum emitters used to generate such light, cold atomic clouds offer unique advantages, since the low temperature strongly suppresses the inhomogeneous broadening that typically affects artificial emitters~\cite{Heshami2016,Gross2017,Shi2018}. Moreover, optically dense cold atomic clouds are of great interest due to their ability to enhance light-matter interactions, enabling strong coupling, directional emission, and collective phenomena such as superradiance~\cite{Dicke1954,Gross1982}. However, high optical thickness causes photons to be scattered multiple times before exiting the cloud, entailing successive Doppler shifts which broaden the spectrum and further reduce the emitted-light coherence time~\cite{Labeyrie2004,Dussaux_2016,Eloy_2018,Ferreira2020,Lassegues_2022}.

\begin{figure}\centering
    \includegraphics[width=\columnwidth]{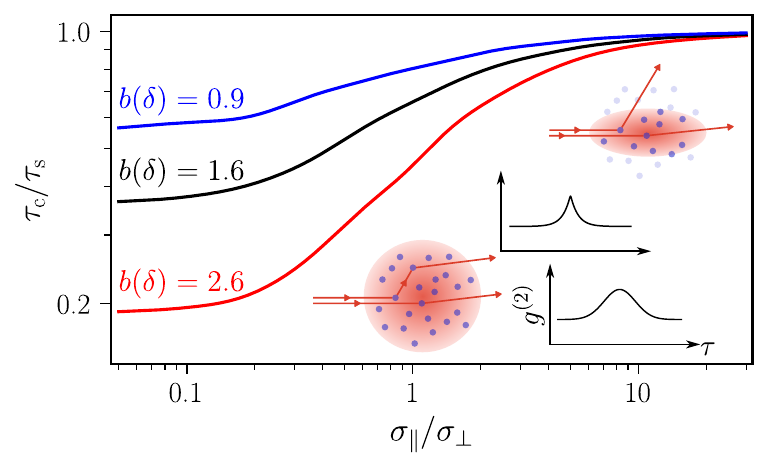}
    \caption{\justifying Coherence time of light scattered by a cold atomic cloud, normalized to the single-scattering coherence time, as a function of cloud aspect ratio for different optical thicknesses $b(\delta)$, obtained from random-walk simulations. Illustrations show typical photon paths in spherical (center) and elongated (top-right) clouds, where lighter atoms are in a dark state and do not scatter. Inset plots show the corresponding $g^{(2)}(\tau)$: a narrow peak (short coherence) for multiple- and a broader Gaussian (long coherence) for single-scattering.}
      \label{Fig:Drawing_MS_3D_cigar}
\end{figure}

In this work, we experimentally demonstrate, through $g^{(2)}$ intensity correlation measurements, that the geometry of an optically dense atomic cloud can be used to control the coherence time of scattered light. By reshaping the cloud from a 3D to a quasi-1D configuration, we reduce the impact of multiple scattering and thereby significantly enhance the temporal coherence of the scattered light, while maintaining the on-axis optical thickness, see Fig.~\ref{Fig:Drawing_MS_3D_cigar}. Furthermore, detection of the scattered light at a small angle reveals two distinct timescales associated with single- and multiple-scattered light,
%by properly choosing the detection angle of the scattered light, we can select singly scattered photons whose coherence time greatly exceeds that of multiply scattered ones,
hence providing a means to discriminate between short- and long-coherence photons. Our approach offers new insights into the role of geometry in shaping the coherence of scattered light in cold atomic systems and enables further control in quantum optics experiments.

\begin{figure*}[t]
    \includegraphics[width=\linewidth]{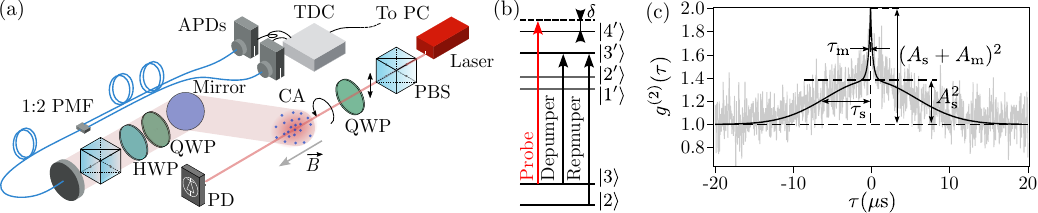}
    \caption{\justifying(a) Experimental setup.  
    A probe laser beam is circularly polarized using a polarizing beam splitter (PBS) and a quarter wave plate (QWP), then sent onto the cold atomic cloud (CA). The transmitted light is detected by a photodetector (PD), and used to determine the on-resonant optical thickness of the cloud. Light from the whole cloud scattered at $\theta=6.3^\circ$ is collected using a single-mode fiber after passing through a QWP, a half-wave plate (HWP), and a PBS to select the relevant polarization (see text). The collected light is then split by a 50:50 fiber beam splitter (1:2 PMF), with the outputs directed to two avalanche photodiodes (APDs). Single-photon counts from each APD are time-tagged using a time-to-digital converter (TDC) and analyzed on a computer (PC). A constant magnetic field $\vec{B}$ along the probe axis sets the quantization axis.
    (b) Relevant atomic levels of the D2 hyperfine transition of $^{85}$Rb. 
    (c) Typical measured intensity correlation function. The gray curve represents experimental data obtained from the 3D cloud with $b(\delta) = 1.56$, while the black curve shows a fit of the data using Eq.~\eqref{eq:fitg2}. The central narrow peak corresponds to multiple scattering, whereas the broader Gaussian feature arises from single-scattered photons.}
    \label{Fig:Setup}
\end{figure*}
\textit{Experimental Setup.} The scattering medium is a cold cloud of $^{85}$Rb atoms produced in a magneto-optical trap (MOT). The cloud has a Gaussian density profile characterized by absorption imaging, yielding the root-mean-square (RMS) radii $\sigma_\parallel$ and $\sigma_\perp$ along the longitudinal and transverse directions, respectively, relative to the probe axis (see~\cite{Eloy_2018,Ortiz_2019} for details). Owing to the symmetry of the setup, $\sigma_x=\sigma_y=\sigma_\perp$, with typical values $\sigma_\perp \approx 1~\mathrm{mm}$ and $\sigma_\parallel \approx 0.8~\mathrm{mm}$, corresponding to an aspect ratio $\sigma_\parallel/\sigma_\perp \simeq 0.8$.
%The cloud has a Gaussian density distribution and is characterized using absorption imaging, which provides the root-mean-square (RMS) sizes $\sigma_\parallel$ and $\sigma_\perp$ along the longitudinal and transverse directions (with respect to the probe axis), see~\cite{Eloy_2018,Ortiz_2019} for details. Due to the symmetry of the setup, $\sigma_x = \sigma_y = \sigma_\perp$, with typical dimensions $\sigma_\perp \approx 1~\mathrm{mm}$ and $\sigma_\parallel \approx 0.8~\mathrm{mm}$, giving an aspect ratio $\sigma_\parallel/\sigma_\perp \simeq 0.8$. In this work, we refer to this nearly isotropic cloud configuration as the \emph{3D cloud}.

To create an elongated cloud geometry, we restrict the volume of atoms which interact with the probe. After the 3D cloud is created, all atoms are optically pumped into the dark hyperfine ground state $|F=2\rangle$ with a short depumper pulse, see Fig.~\ref{Fig:Setup}(b) for the relevant level scheme. A repumper beam, resonant with the $|F=2\rangle \to |F=3^\prime\rangle$ transition and coupled into the same fiber as the probe, transfers {\it only the atoms within the probe mode} back into the $|F=3\rangle$ ground state (dark-colored circles in the right inset of Fig.~\ref{Fig:Drawing_MS_3D_cigar}). The repumper and probe beams have 100\,$\mu$m waist, small compared to the 3D cloud size. Thus, only the atoms within the elongated region illuminated by the repumper are in a state to interact with the probe light. This configuration effectively creates a {\it quasi-1D} atomic cloud, with its long axis along the probe propagation direction. The typical dimensions are $\sigma_\perp\approx120~\mu$m and $\sigma_\parallel\approx480~\mu$m, yielding an aspect ratio of $\sigma_\parallel/\sigma_\perp\approx 4$.

After preparation, the 3D or quasi-1D cloud is illuminated by a circularly polarized probe [see Fig.~\ref{Fig:Setup}(a)] tuned near the $|F=3\rangle \to |F=4^\prime\rangle$ $D_2$ transition. The probe first optically pumps the atoms into the $m_F = 3$ Zeeman sublevel, with a magnetic field along the probe axis lifting the Zeeman degeneracy and stabilizing the pumping process. Once optically pumped, the circularly polarized probe drives a nearly closed $|F=3,m_F=3\rangle \to |F'=4,m_F'=4\rangle$ cycling transition, so the atoms behave as effective two-level systems.
%Once optically pumped, the circular polarization of the light ensures selective excitation toward the $m_F'=4$ sublevel, so the atoms, in a nearly closed cycling transition, effectively behave as two-level systems. 
Thus, in the forward direction, the first scattering event preserves probe polarization.

For a given cloud geometry, the amount of multiply scattered light is set by the optical thickness of the cloud along the probe axis, 
%a parameter that allows us to tune the multiple scattering contribution is the optical thickness of the cloud along the probe axis, 
\begin{equation}
    b(\delta) = \dfrac{b_0}{1 + 4\delta^2/\Gamma^2},
\end{equation}
where $\delta$ is the detuning of the probe from the atomic transition frequency, $\Gamma$ the natural linewidth, and %$b_0=\sigma_\mathrm{sc}\int\rho(0,0,z)~\mathrm{d}z$ the on-resonance optical thickness along the probe axis $z$, with $\sigma_\mathrm{sc}=3\lambda^2/2\pi$ the atomic scattering cross-section, $\lambda$ the wavelength and $\rho$ the spatial density of atoms in the $|F=3\rangle$ state.We determine $b_0$ from off-resonant transmission measurements fitted with the Beer–Lambert law, $\mathcal{T}(\delta)=\exp[-b(\delta)]$, and obtain $b(\delta)$ for a given detuning.
$b_0$ the on-resonance optical thickness which is determined from off-resonant transmission measurements fitted with the Beer–Lambert law, $\mathcal{T}(\delta)=\exp[-b(\delta)]$. By varying $\delta$, we directly tune $b(\delta)$. The saturation parameter of the probe is fixed at $s(\delta)=0.1$, ensuring negligible inelastic scattering~\cite{Ortiz_2019,Lassegues_2023}.

 %The probe intensity is adjusted accordingly, keeping the saturation parameter fixed at $s(\delta) = 0.1$, ensuring a negligible contribution from inelastic scattering~\cite{Ortiz_2019,Lassegues_2023}.

Scattered probe light is collected at $\theta = 6.3 \pm 1^\circ$ from the probe axis. The detection polarization optics (QWP, HWP, PBS) are aligned to transmit the probe polarization, maximizing the collection of singly scattered photons, which largely preserve polarization (due to a small detection angle), while partially transmitting the multiply scattered ones, as they are more depolarized. The collected light is then coupled into a single-mode fiber and split between two avalanche photodiodes (APDs), forming a Hanbury Brown–Twiss (HBT) setup~\cite{HBT:1956a},  as detailed in Fig.~\ref{Fig:Setup}(a). A time-to-digital converter (TDC) records the photon arrival times from both APDs, from which we construct coincidence histograms between the two APDs as a function of delay $\tau$~\cite{Lassegues_2023}. These yield the second-order intensity correlation function~\cite{Eloy_2018}, defined classically as
\begin{equation}
g^{(2)}(\tau) = \frac{\langle I(t) I(t+\tau)\rangle}{\langle I(t)\rangle^2},
\end{equation}
%The output signals are time-tagged by a time-to-digital converter (TDC), which records the arrival times of individual photons~\cite{Lassegues_2023} and allows us to build the histogram of coincidences between the two detectors as function of the delay $\tau$. Normalizing these histograms provides the experimental second-order intensity correlation function~\cite{Eloy_2018}, defined classically as
%\begin{equation}
%    g^{(2)}(\tau) = \frac{\langle I(t)I(t + \tau) %\rangle}{\langle I(t) \rangle^2},
%\end{equation}
%where $\langle\cdot\rangle$ denotes a time average, and $I$ the classical intensity dynamics. In practice, $I(t)$ represents the photon detection events registered by the APDs, and $\langle\cdot\rangle$ ensemble averages over many realizations of the experiment (assuming ergodicity). 
where in practice, $I(t)$ represents the photon detection events registered by the APDs, and $\langle\cdot\rangle$ ensemble averages over many realizations of the experiment. 

\textit{Temporal coherence of single and multiple scattering.} An example of $g^{(2)}(\tau)$ measured from the 3D cloud is shown in Fig.~\ref{Fig:Setup}(c). Two distinct features are observed, corresponding to single- and multiple-scattering contributions with notably different coherence times.
%, where two structures with different coherence times are observed, which can be associated with single- and multiple-scattering processes.
For single scattering, the Gaussian spectrum arising from the Maxwell-Boltzmann distribution of velocities leads to an electric field autocorrelation~\cite{Loudon:book}
\begin{equation}
    g^{(1)}_\mathrm{s}(\tau) = \frac{\langle E^*(t)E(t + \tau) \rangle}{\langle I(t) \rangle} = \exp\left(-\frac{\tau^2}{4\tau_\mathrm{s}^2}\right),
\end{equation}
characterized by its coherence time $\tau_\mathrm{s}$ given by 
\begin{equation}
    \tau_\mathrm{s}^{-1} = 2k_\mathrm{L} \sqrt{k_\mathrm{B} T (1 - \cos\theta)/m},
    \label{eq:tausingle}
\end{equation}
with $k_\mathrm{L}$ the probe wavenumber, $m$ the atomic mass, $k_\mathrm{B}$ the Boltzmann constant and $T$ the cloud temperature.
Furthermore, for chaotic (also called thermal) light sources, such as light scattered by cold atoms, this temporal coherence translates directly into intensity correlations through the Siegert relation, $g^{(2)}(\tau)=1+|g^{(1)}(\tau)|^2$~\cite{Ferreira2020,Lassegues_2022}.

In contrast, the multiple-scattering field correlation function $g^{(1)}_\mathrm{m}(\tau)$ strongly depends on $b(\delta)$, as multiple scattering events reduce the coherence time of the light~\cite{Eloy_2018,Dussaux_2016}. 
In particular, while single scattering exhibits a long coherence time in the forward direction due to the $(1-\cos\theta)$ prefactor of the Doppler effect, multiply scattered light does not benefit from this directionality effect.
Hence, at small angle $\theta \approx 6.3^\circ$ used in our setup, single- and multiple-scattering contributions appear on clearly separated timescales in $g^{(2)}(\tau)$, enabling their discrimination.

%the small angle $\theta\approx 6.3^\circ$ used in our setup leads to two very different timescales for single and multiple scattering, which in turn allows one to discriminate each contribution in the $g^{(2)}(\tau)$.

Decomposing the intensity into single- and multiple scattering components, $I_\mathrm{s}$ and $I_\mathrm{m}$ respectively, $g^{(2)}(\tau)$ can be written as a weighted combination~\cite{SM}: 
%\begin{equation}
%\begin{aligned}
%    g^{(2)}(\tau) = 1 + C \left[ \frac{I_\mathrm{s} \left| g^{(1)}_\mathrm{s}(\tau) \right| + I_\mathrm{m}\left| g^{(1)}_\mathrm{m}(\tau) \right|}{I_\mathrm{tot}}\right]^2,
%\end{aligned}
%\end{equation}
\begin{equation}
\begin{aligned}
    g^{(2)}(\tau) = 1 +  \left[ \frac{I_\mathrm{s} \left| g^{(1)}_\mathrm{s}(\tau) \right| + I_\mathrm{m}\left| g^{(1)}_\mathrm{m}(\tau) \right|}{I_\mathrm{tot}}\right]^2,
\end{aligned}
\end{equation}
with $I_\mathrm{tot} = I_\mathrm{s} + I_\mathrm{m}$, the total collected scattered light. 
%The parameter $C=g^{(2)}(0)-1$ accounts for the reduced contrast in the experiment, mainly due to APDs dark counts, with typical values of $C\approx 0.8$ instead of $C = 1$ for a perfectly bunched, thermal source. 

%\ac{For the measured correlation function, we first determine the parameter $C=g^{(2)}(0)-1$ which accounts for the reduced contrast in the experiment, mainly due to APDs dark counts, with typical values of $C\approx 0.8$ instead of $C = 1$ for a perfectly bunched, thermal source. The correlation function is then normalized as $[g^{(2)}(\tau)-1]/C$, and to extract the characteristic times and relative weights we fit the data with}
To extract the characteristic times and relative weights, we fit the data with
\begin{equation}
    g^{(2)}(\tau) = 1 + \left( A_\mathrm{s} \, \mathrm{e}^{-\tau^2 / 4\tau_\mathrm{s}^2} + A_\mathrm{m} \, \mathrm{e}^{-|\tau| / \sqrt{2\pi}\tau_\mathrm{m}} \right)^2.
    \label{eq:fitg2}
\end{equation}
Here, $\tau_\mathrm{s}$ and $\tau_\mathrm{m}$ denote the coherence times of single and multiple scattering, and $A_\mathrm{s}$, $A_\mathrm{m}$ their amplitudes satisfying %$A_\mathrm{m}/A_\mathrm{s}=I_\mathrm{m}/I_\mathrm{s}$ and $(A_\mathrm{s}+A_\mathrm{m})^2=C$. 
$A_\mathrm{s}=I_\mathrm{s}/I_\mathrm{tot}$ and $A_\mathrm{m}=I_\mathrm{m}/I_\mathrm{tot}$
The multiple-scattering term is modeled by an exponential decay, a good approximation for cold atomic clouds~\cite{Eloy_2018}, as shown in Fig.~\ref{Fig:Setup}(c). Note that, before fitting, all experimental second-order correlation functions are 
renormalized such that $g^{(2)}(0)=2$, compensating for the reduced contrast, mainly
due to APDs dark counts. In this work, whenever we refer to measured
$g^{(2)}(\tau)$, it is understood that the curves are already renormalized 
in this way, such that $A_\mathrm{s}+A_\mathrm{m}=1$.

\textit{Control of the temporal coherence through geometry.} We now demonstrate how the cloud geometry controls the relative weight of single and multiple scattering, and in turn the coherence time of the scattered light. 
%On the one hand, fitting the experimental data using equation~\eqref{eq:fitg2} provides the amplitudes of single and multiple scattering, $A_\mathrm{s}$ and $A_\mathrm{m}$. On the other hand the coherence time $\tau_\mathrm{c}$ is computed directly as the integral of the excess intensity correlation~\cite{Loudon:book}:

From fits to Eq.\eqref{eq:fitg2} we extract the amplitudes $A_\mathrm{s}$ and $A_\mathrm{m}$ associated with single and multiple scattering contribution. Independently, we define the total coherence time $\tau_\mathrm{c}$ as the integral of the excess intensity correlation\cite{Loudon:book},
\begin{equation}
\tau_\mathrm{c} =\frac{1}{\sqrt{2\pi}} \int \left[g^{(2)}(\tau) - 1\right]\,\mathrm{d}\tau,\label{eq:defg2}
\end{equation}
%a definition which encompasses both contributions in Eq.~\eqref{eq:fitg2}.
which incorporates both scattering terms in Eq.~\eqref{eq:fitg2}.

%\begin{equation}
%\tau_\mathrm{c} =\frac{1}{\sqrt{2\pi}} \int \frac{\left[g^{(2)}(\tau) - 1\right]}{C}\,\mathrm{d}\tau,\label{eq:defg2}
%\end{equation}

\begin{figure*}[t]
    \centering
    \begin{subfigure}[b]{0.49\textwidth}
        \includegraphics[width=\linewidth]{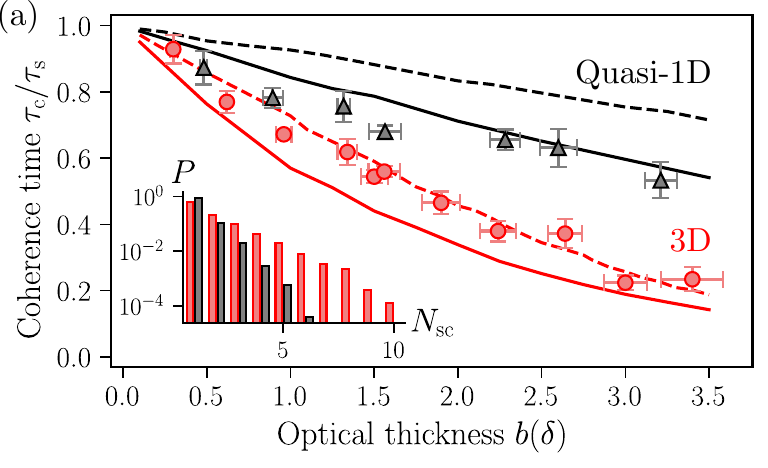}
%     \label{Fig:fig3a}
    \end{subfigure}
    \begin{subfigure}[b]{0.49\textwidth}
        \includegraphics[width=\linewidth]{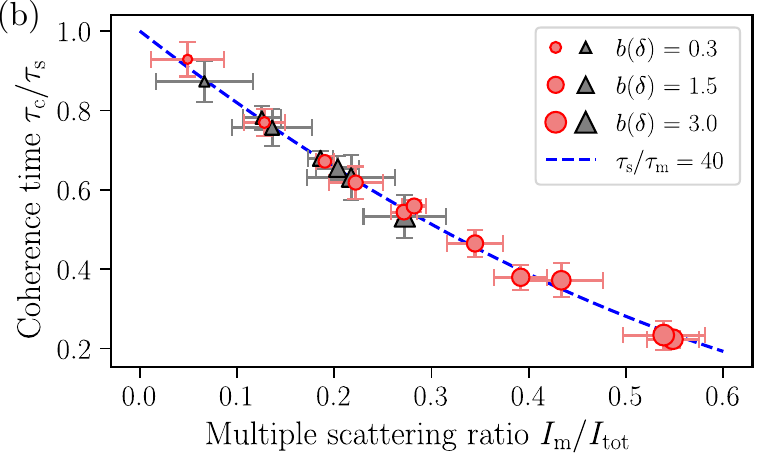}
%        \label{Fig:fig3b}
    \end{subfigure}
    \caption{\justifying (a) Coherence time $\tau_\textrm{c}$, normalized by the single-scattering coherence time $\tau_\textrm{s}$, as a function of the optical thickness $b(\delta)$. Symbols: experimental data for 3D (red circles) and quasi-1D (black triangles) cold atomic cloud. Solid (dashed) curves: random-walk simulations including polarization at all events (only at the first event, with subsequent scatterings treated as depolarized), with polarization filtering at detection; see main text. Inset: Probability distribution of the number of scattering events $N_\mathrm{sc}$ for $b(\delta) = 1.6$. (b) Coherence time as a function of the relative weight of multiple scattering, $I_\mathrm{m}/I_\mathrm{tot}$. Symbols correspond to the same datasets as in (a), with marker size scaling with $b(\delta)$ (see legend for examples). The dashed line corresponds to Eq.~\eqref{eq:ratiotsm} with $\tau_\mathrm{s}/\tau_\mathrm{m} = 40$.
    }
    \label{Fig:fig3}
\end{figure*}

As shown in Fig.~\ref{Fig:fig3}(a), the coherence time decreases with increasing optical thickness
%$b(\delta)$, a manifestation of the increasing relative weight of multiple scattering, i.e. the increasing number of scattering events undergone by a photon and the associated increase of the inhomogeneous Doppler broadening.
This trend reflects the increasing contribution of multiple scattering: as photons undergo more scattering events, the associated increase of Doppler broadening reduces coherence.
In contrast, the single-scattering coherence time $\tau_\mathrm{s}$ depends only on the temperature and the detection angle, not on the cloud geometry, and therefore provides a reference timescale -- it is hereafter used as a normalization factor for $\tau_\mathrm{c}$.

More interestingly, quasi-1D clouds present systematically longer coherence times than the 3D ones, for a given optical thickness (black triangles vs. red circles in Fig.~\ref{Fig:fig3}(a)). This can be understood from the fact that in a quasi-1D geometry photons are more likely to exit the sample after a single scattering event, as illustrated by the clouds in the insets of Fig.~\ref{Fig:Drawing_MS_3D_cigar}. Although the two configurations may have the same on-axis optical thickness, the reduced transverse size of the quasi-1D cloud lowers the probability of subsequent scattering events, leading to a smaller contribution from multiple scattering (Fig.~\ref{Fig:Angular_dist} in ~\cite{SM}). This is quantified by the probability distribution $P$ of the number of scattering events $N_\textrm{sc}$ in the inset of Fig.~\ref{Fig:fig3}(a), where the quasi-1D geometry (in black) exhibits a much faster decay with $N_\textrm{sc}$ than the 3D one (in red) for same on-axis optical thickness. This result shows the role of the geometry in preserving the coherence of the scattered light of optically thick clouds.

The role of the relative contribution of multiply scattered light can be better understood using Eq.~\eqref{eq:fitg2}. In the limit of a large ratio $\tau_\mathrm{s}/\tau_\mathrm{m}$, which in our case is of the order of $\sim 40$, 
%definition given by Eq.~\eqref{eq:defg2} for the coherence time leads to~\cite{SM}
the definition of $\tau_\mathrm{c}$ [Eq.\eqref{eq:defg2}] yields~\cite{SM}
\begin{equation}
    \frac{\tau_\mathrm{c}}{\tau_\mathrm{s}} = \left(1-\frac{I_\mathrm{m}}{I_\mathrm{tot}}\right)^2 + \frac{I_\mathrm{m}}{I_\mathrm{tot}}\left(4-3\frac{I_\mathrm{m}}{I_\mathrm{tot}}\right)\frac{\tau_\mathrm{m}}{\tau_\mathrm{s}}.\label{eq:ratiotsm}
\end{equation}
In the expression above, for $\tau_\mathrm{m}\ll \tau_\mathrm{s}$ and  $I_\mathrm{m}/I_\mathrm{tot}$ not close to 1, as is the case in our setup, 
the ratio $\tau_\mathrm{c}/\tau_\mathrm{s}$ is governed predominantly by $I_\mathrm{m}/I_\mathrm{tot}$. This is shown in Fig.~\ref{Fig:fig3}(b), where the experimental data for atomic clouds of different geometries collapse on a single one. The dashed line shows Eq.~\eqref{eq:ratiotsm} for $\tau_\mathrm{s}/\tau_\mathrm{m} = 40$, which is a typical value obtained from fits of the measured $g^{(2)}(\tau)$ [see Fig.~\ref{Fig:Setup}(c)].

Further insights into the multiple scattering process, and in particular the resulting depolarization, can be obtained from photon random walk (RW) simulations~\cite{Chabe2014,Sokolov2019,Kemp2020}, where the polarization is treated by assuming that the atoms act as classical dipoles, or equivalently as atoms undergoing a $|J=0\rangle \to |J'=1\rangle$ transition.
%The detection setup is aligned to the polarization of the incident probe, thereby maximizing the collection of singly scattered photons at small detection angle in the collection fiber mode, while only a fraction of multiply scattered photons is collected due to their more randomized polarization. To account for this polarization effect, 
To account for the depolarization and polarization filtering used in our detection setup, the RW simulations are performed considering two models~\cite{SM}: {\it polarization-tracking} model which includes the full polarization dependence at each scattering event (solid lines in Fig.~\ref{Fig:fig3}(a)), or {\it depolarized} model which considers polarization only at the first scattering, with subsequent scatterings treated as fully depolarized, implemented by a 50\% reduction of the number of detected multiply scattered photons to mimic polarization filtering (dashed lines in Fig.~\ref{Fig:fig3}(a)). 

For the 3D cloud, the experimental data agree well with the depolarized model. This is consistent with the larger number of scattering events, which reduces the degree of polarization. %Two extra experimental factors contribute to the stronger depolarization: (i) the $|F=3\rangle \to |F=4^\prime\rangle$ transition of $^{85}$Rb involves many Zeeman sublevels with different Clebsch–Gordan coefficients, unlike the closed cycling transition used in RW simulations, so re-absorption and re-emission can occur on different sublevels, enhancing the polarization mixing of multiply scattered photons, and (ii) partial optical pumping where only the central region of the 3D cloud is optically pumped by the probe beam into the stretched state (since the probe beam is much smaller than the 3D cloud size), whereas the atoms in the outer region remain distributed among different Zeeman sublevels. Both effects contribute to faster depolarization and make multiple scattering effectively fully depolarized, consistent with the simplified model of considering polarization only for the first scattering event. 
Moreover, in the experiment, additional depolarization arises from (i) the complex Zeeman structure of the $|F=3\rangle \to |F'=4\rangle$ transition and (ii) partial optical pumping, since only the central region of the cloud is driven into the stretched state (see~\cite{SM} for more details). These effects lead to a faster depolarization and make multiple scattering effectively fully depolarized, justifying the simplified model where polarization is preserved only at the first scattering. In contrast, in the quasi-1D cloud the whole volume is efficiently pumped into the stretched state, and the reduced number of scattering events further preserves polarization. Consequently, the experimental data agree more closely with the polarization-tracking model.

%In contrast, for the quasi-1D cloud, the entire cloud volume is optically pumped into the stretched state. Combined with the reduced number of scattering events in this elongated geometry, this leads to a better preserved polarization and, as a result, the experimental data lie closer to the partial-depolarized model.

Finally, we use RW simulations to study systematically the transition from 3D to quasi-1D by varying the longitudinal size $\sigma_\parallel$ at fixed transverse size $\sigma_\perp$. The results, shown in Fig.~\ref{Fig:Drawing_MS_3D_cigar}, demonstrate that for a constant on-axis optical thickness $b(\delta)$, the coherence time $\tau_\mathrm{c}$ increases as the cloud is elongated i.e., as the aspect ratio $\sigma_\parallel/\sigma_\perp$ grows, approaching the single-scattering limit $\tau_\mathrm{s}$ in the $\sigma_\parallel/\sigma_\perp \gg 1$ limit. 
%Moreover, despite larger optical thicknesses are characterized by lower coherence times (see $b(\delta)=2.6$ red curve in Fig.~\ref{Fig:Drawing_MS_3D_cigar}) as the Doppler broadening becomes stronger, the quasi-1D regime still allows to recover the $\tau_\mathrm{c}\to\tau_\mathrm{s}$ limit. This trend is consistent across the optical thicknesses explored, and it results from the enhanced suppression of the multiple scattering as the cloud aspect ratio increases~\cite{SM}, providing a robust means to control the temporal coherence. 
Even at larger $b(\delta)$, where Doppler broadening reduces coherence (e.g. $b(\delta)=2.6$, red curve in Fig.~\ref{Fig:Drawing_MS_3D_cigar}), the quasi-1D geometry still drives $\tau_\mathrm{c}\to\tau_\mathrm{s}$. This trend, observed across all explored $b(\delta$), reflects the enhanced suppression of multiple scattering with increasing aspect ratio, providing a robust means for controlling temporal coherence in optically dense clouds.

\textit{ Conclusions \& Perspectives.} Using intensity correlation measurements performed on cold atomic clouds of different geometries and optical thicknesses, we have experimentally demonstrated how to increase the coherence time of the scattered light by reaching the quasi-1D limit. Using a repumper to shape an effective elongated cloud out of a spherical one, the coherence time can be increased to the single-scattering limit, where the coherence is set by the angle of observation and the temperature only.

Our results highlight the potential of manipulating the dimensionality of atomic clouds to engineer the coherence properties of light. This approach could be particularly useful for the observation of non-classical photon statistics in large systems~\cite{Prasad_2020,Cordier2023,Bojer2024,Singh2025}, such as antibunching in the light transmitted by a cold atomic cloud, or to suppress decoherence in light-matter interfaces. Future work could extend this approach to frequency-resolved or cross-polarization correlation measurements, offering new opportunities for the generation of quantum light in atomic systems.

\begin{acknowledgments}
We wish to acknowledge Lucas Pache from Department of Physics, Humboldt-Universit\"at zu Berlin, Germany, for the experimental implementation of the quasi-1D cloud. M.\,H., R.\,B., R.\,C.\,T. and R.\,K. acknowledge the ANR-FAPESP grant (ANR19-CE47-0014-01).  M.\,H. and R.\,B. has been supported by the the UCA J.E.D.I. Investments (ANR-15-IDEX-01). R.K. received support from the European project ANDLICA, ERC Advanced Grant No. 832219.
R.\,K., R.\,B., R.~C.~T. and M.\,H. received support from STIC-AmSud (Ph879-17/CAPES 88887.521971/2020-00) and CAPES-COFECUB (Ph 997/23, CAPES 88887.711967/2022-00). R.~B., R.~C.~T. and M.~A.~F.~B.~acknowledge the support from the S\~ao Paulo Research Foundation (FAPESP, Grants Nos. 2018/23873-3, 2019/13143-0, 2021/02673-9, 2022/00209-6, 2023/15457-8 and 2023/03300-7) and from the National Council for Scientific and Technological Development (Grant Nos.\,313632/2023-5 and 403653/2024-0). A.C. acknowledges funding from the Brazilian funding agency FAPESP via Grant Nos. 2023/00866-0 and 2024/13462-7. J. C. C. C. acknowledges support from CAPES-COFECUB (88887.898622/2023-00), Office of Naval Research (ONR Grant No. N62909-23-1-2014), Financiadora de Estudos e Projetos (MCTI/FINEP/FINDCT 
CENTROS TEMATICOS 2023 Grant No 1020/24) and Fundação de Amparo à Pesquisa do Estado de São Paulo (FAPESP Grant No. 2021/06535-0).
\end{acknowledgments}

\bibliography{Biblio}{}% Produces the bibliography via BibTeX.

\onecolumngrid
\newpage

% \pagebreak

% \clearpage

%####################################################################%
%##############      SUPPLEMENTAL MATERIAL     ######################%
%####################################################################%
%%%%%%%%%% Prefix a "S" to all equations, figures, tables and reset the counter %%%%%%%%%%% 
\setcounter{equation}{0}
\setcounter{figure}{0}
\setcounter{table}{0}
\setcounter{page}{1}
% \makeatletter
\renewcommand{\theequation}{S\arabic{equation}}
\renewcommand{\thefigure}{S\arabic{figure}}
%\renewcommand{\bibnumfmt}[1]{[S#1]}
%ß\renewcommand{\citenumfont}[1]{S#1}

\begin{center}
%%%%%%%%% ABSTRACT TITLE
{\large{ {\bf Supplemental Material for: \\ Dimensional Control of the Coherence Time of Scattered Light in Cold Atom Clouds}}}

%%%%%%%%% ABSTRACT AUTHORS
\vskip0.5\baselineskip{Ana Cipris,$^{1,3}$ Mateus A. F. Biscassi,$^{1,2}$ J. C. C. Capella,$^{1,4}$ Martial Morisse,$^{1}$ Hani Naim,$^{1}$ Hugo Sedlacek,$^{1}$ Apoorav Singh Deo,$^{1}$ Stephan Asselie,$^{1}$ Robin Kaiser,$^{1}$ Raul Celistrino Teixeira,$^{2}$ R. Bachelard$^{2}$ and Mathilde Hugbart,$^{1}$}

%%%%%%%%% AFFILIATION^{1}
\vskip0.5\baselineskip{ {\it $^{1}$Universit\'e C\^ote d'Azur, CNRS, INPHYNI, France\\
$^{2}$Departamento de F\'isica, Universidade Federal de S\~ao Carlos,\\ Rodovia Washington Lu\'is, km 235—SP-310, 13565-905 S\~ao Carlos, SP, Brazil \\ $^{3}$Instituto de Física de São Carlos, Universidade de São Paulo, São Carlos SP 13566-970, Brazil \\ $^{4}$Departamento de F\'{\i}sica, Universidade Federal de Pernambuco, 50670-901 Recife, Pernambuco, Brazil
}}

\end{center}

\appendix

\setcounter{equation}{0}
\setcounter{figure}{0}
\setcounter{table}{0}
\setcounter{page}{1}

\twocolumngrid

\section{Total $g^{(2)}(\tau)$, multiple-scattering ratio and coherence time}
To extract the ratio of multiple scattering versus total scattering $I_\mathrm{m}/I_\mathrm{tot}$, we consider the total scattered power spectrum $S(\nu)$, which is the sum of the single- and multiple-scattering contributions $S_\mathrm{s}$ and $S_\mathrm{m}$, respectively:
\begin{equation}
    S(\nu) = S_\mathrm{s}(\nu) + S_\mathrm{m}(\nu).
\end{equation}
The corresponding scattered intensities are proportional to the inverse Fourier transforms of the respective spectra evaluated at zero delay:
\begin{eqnarray}
    I_\mathrm{s} &\propto& \mathcal{F}^{-1}\{S_\mathrm{s}(\nu)\}|_{\tau=0}\\% \int S_\mathrm{s}(\nu) d\nu\\ % = g^{(1)}_\mathrm{s}(0),\\
    I_\mathrm{m} &\propto& \mathcal{F}^{-1}\{S_\mathrm{m}(\nu)\}|_{\tau=0}. %\int S_\mathrm{m}(\nu) d\nu. % = g^{(1)}_\mathrm{m}(0),
\end{eqnarray}
According to the Wiener-Khinchine theorem~\cite{Wiener1930}, the first order correlation function is given by the normalized inverse Fourier transform of the total power spectrum:
\begin{eqnarray}
g^{(1)}(\tau) &=& \frac{\mathcal{F}^{-1}\{S(\nu)\}}{\mathcal{F}^{-1}\{S(\nu)\}|_{\tau=0}},\\
%g^{(1)}(\tau) &=& \frac{\mathrm{FFT}\left[S(\nu) \right]}{\mathrm{FFT}\left[ S(\nu) \right](\tau = 0)},\\
%&=& \frac{\mathrm{FFT}\left[S(\nu) \right]}{\int S(\nu) d\nu},\\
%&=& \frac{\mathrm{FFT}\left[S(\nu) \right]}{I_\mathrm{s} + I_\mathrm{m}},\\
&=& \frac{I_\mathrm{s}}{I_\mathrm{tot}} g_\mathrm{s}^{(1)}(\tau) + \frac{I_\mathrm{m}}{I_\mathrm{tot}} g_\mathrm{m}^{(1)}(\tau),
\end{eqnarray}
where $I_\mathrm{tot}=I_\mathrm{s} + I_\mathrm{m}$. Assuming the scattered light exhibits Gaussian statistics, as expected from light emitted by a cold atomic ensemble~\cite{Eloy_2018,Lassegues_2022,Lassegues_2023,Morisse_2024}, the intensity correlation function is related to the field correlation via the Siegert relation~\cite{Siegert:1943}:
\begin{eqnarray}
g^{(2)}(\tau) &=& 1+ |g^{(1)}(\tau)|^2, \label{Eq:Siegert}\\
%&=& 1+C \left\lvert\frac{I_\mathrm{s}}{I_\mathrm{s} + I_\mathrm{m}} g_\mathrm{s}^{(1)}(\tau) + \frac{I_\mathrm{m}}{I_\mathrm{s} + I_\mathrm{m}} g_\mathrm{m}^{(1)}(\tau)\right\rvert^2,\\
&=& 1 +  \left[ \frac{I_\mathrm{s} \left| g^{(1)}_\mathrm{s}(\tau) \right| + I_\mathrm{m}\left| g^{(1)}_\mathrm{m}(\tau) \right|}{I_\mathrm{tot}}\right]^2,
\end{eqnarray}
providing an expression for the second-order correlation function that accounts for both single and multiple scattering contributions.

%To quantify the relative contribution of multiple scattering, we first determine the contrast of the intensity correlation function, defined as $C = g^{(2)}(\tau=0)-1$. \ac{The measured correlation function is then normalized as $[g^{(2)}(\tau)-1]/C$, and subsequently fitted using Eq.\,\eqref{eq:fitg2}.}
The measured $g^{(2)}(\tau)$ is fitted using Eq.\,\eqref{eq:fitg2}. 
From the fit, the amplitude associated with the single and multiple-scattering components, $A_\mathrm{s}$ and $A_\mathrm{m}$, are extracted. The corresponding intensity ratios are given by:
\begin{eqnarray}
    \frac{I_\mathrm{s}}{I_\mathrm{tot}} &=& A_\mathrm{s},\label{eq:Is} \\
    \frac{I_\mathrm{m}}{I_\mathrm{tot}} &=& 1-\frac{I_\mathrm{s}}{I_\mathrm{tot}} = 1 - A_\mathrm{s}=A_\mathrm{m}.\nonumber
\end{eqnarray}
These expressions provide a direct estimation of the single- and multiple-scattering contributions from the measured correlation function and fit parameters.

Finally, the ansatz~\eqref{eq:fitg2} for the coherence time, combined with definition~\eqref{eq:defg2}, results in
\begin{equation}
\begin{aligned}
    \tau_c=&A_\mathrm{s}^2 \tau_\mathrm{s} + A_\mathrm{m}^2 \tau_\mathrm{m} \\ &+ 2\sqrt{2}A_\mathrm{m} A_\mathrm{s} \tau_\mathrm{s} e^{-\frac{\tau_\mathrm{s}^2}{2\pi\tau_\mathrm{m}^2}}\textrm{erfc}\left(\frac{\tau_\mathrm{s}}{\sqrt{2\pi}\tau_\mathrm{m}}\right),
\end{aligned}
\end{equation}
with $\textrm{erfc}$ the complementary error function. In the $\tau_\mathrm{s}/\tau_\mathrm{m}\gg 1$ limit (the ratio is of several tens in our setup), using the fact that $e^{x^2}\textrm{erfc}(x)\approx 1/(\sqrt{\pi}x)$ for large $x$, one can approximate the coherence time by
\begin{equation}
    \tau_c=A_\mathrm{s}^2 \tau_\mathrm{s} + A_\mathrm{m}^2 \tau_\mathrm{m} + 4A_\mathrm{m} A_\mathrm{s} \tau_\mathrm{m}.
\end{equation}
Normalizing by the single-scattering coherence time $\tau_\mathrm{s}$, and using relations for $A_\mathrm{s}$ and $A_\mathrm{m}$ from Eq.~\eqref{eq:Is}, we obtain
\begin{align}
\frac{\tau_c}{\tau_s} &= \left(1 - \frac{I_m}{I_{\mathrm{tot}}}\right)^2 \\
&+ \left[\left(\frac{I_m}{I_{\mathrm{tot}}}\right)^2 
+ 4\frac{I_m}{I_{\mathrm{tot}}}\left(1 - \frac{I_m}{I_{\mathrm{tot}}}\right)\right]\frac{\tau_m}{\tau_s},
\end{align}
which simplifies to
\begin{equation}
\frac{\tau_c}{\tau_s} = \left(1 - \frac{I_m}{I_{\mathrm{tot}}}\right)^2 
+ \frac{I_m}{I_{\mathrm{tot}}}\left(4 - 3\frac{I_m}{I_{\mathrm{tot}}}\right)\frac{\tau_m}{\tau_s}.
\end{equation}

\section{Polarization in the random walk simulations}

In order to include polarization effects in the RW simulations, we adopt the Stokes--Mueller formalism, which provides a convenient framework to describe how the polarization state of light evolves during successive scattering events and, through the intensity component of the Stokes vector, the angular probability distribution of scattered photons \cite{Bartel_00, Ramella-Roman_05}. In this approach, the polarization state of each photon is represented by the Stokes vector
 
\begin{equation}
\mathbf{S} = (S_0,\, S_1,\, S_2,\, S_3)^T ,
\end{equation}
where $S_0$ denotes the total intensity, $S_1$ and $S_2$ correspond to linear polarization components, and $S_3$ quantifies the circular polarization. The Stokes vector is always defined with respect to a reference polarization basis. This basis is tied to the scattering plane which is defined by the ingoing and outgoing photon wavevectors. This ensures that the components of $\mathbf{S}$ directly encode the polarization relative to the geometry of each scattering event.

%The action of a scattering event on the polarization state is described by a real $4\times 4$ Mueller matrix $M$, such that the outgoing polarization state is
%\begin{equation}
%\mathbf{S}' = M\,\mathbf{S}.
%\end{equation}

The transformation of the Stokes vector upon each scattering event is given by
\begin{equation}
\mathbf{S}' = M(\theta)\,R(\phi)\,\mathbf{S},
\label{eq:Stokes_update}
\end{equation}
where $\theta$ is the scattering angle between the ingoing and outgoing photon propagation directions, and $\phi$ is the azimuthal angle defining the orientation of the scattering plane around the ingoing photon propagation axis. Moreover, $M(\theta)$ is the Mueller matrix which encodes how the polarization state is modified by the scattering event, while $R(\phi)$ is the rotation matrix that rotates the Stokes vector, i.e. adjusts the polarization reference axis to the new scattering plane.

\begin{figure*}\centering
    \includegraphics[width=\textwidth]{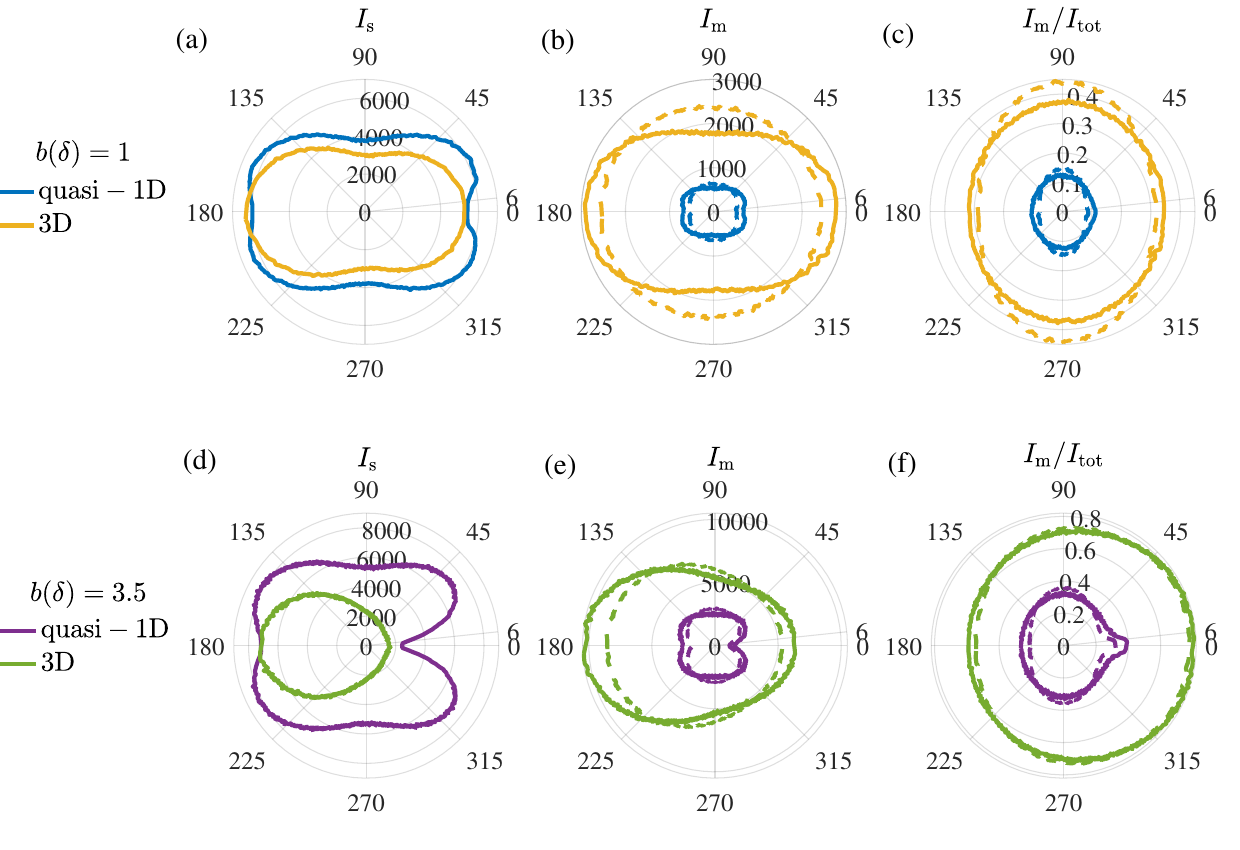}
	\caption{\justifying Angular distribution of singly (a,d) and multiply (b,e) scattered photons for optical thicknesses 
$b(\delta)=1$ (a--c) and $b(\delta)=3.5$ (d--f) in quasi-1D and 3D cold atom configuration, obtained from RW simulations. 
Panels (c,f) show the 
multiple-scattering ratio $I_\mathrm{m}/I_\mathrm{tot}$, where $I_\mathrm{tot}=I_\mathrm{s}+I_\mathrm{m}$. The solid curves correspond to the model where polarization is considered at each scattering event (polarization-tracking model), while dashed lines represent the case where polarization is taken into account only for the first scattering event (depolarized model). For these angular distribution plots, the polarization filtering has not been included.}
      \label{Fig:Angular_dist}
\end{figure*}

Treating the atoms as classical dipoles, or equivalently as atoms undergoing a closed cycling transition $|J=0\rangle \to |J'=1\rangle$ transition, the Mueller matrix reads

\begin{equation}
M(\theta) = \frac{3}{4}
\begin{pmatrix}
1+\cos^2\theta & \sin^2\theta & 0 & 0 \\
\sin^2\theta & 1+\cos^2\theta & 0 & 0 \\
0 & 0 & 2\cos\theta & 0 \\
0 & 0 & 0 & 2\cos\theta
\end{pmatrix},
\end{equation}
and the rotation matrix is
\begin{equation}
R(\phi) =
\begin{pmatrix}
1 & 0 & 0 & 0 \\
0 & \cos 2\phi & \sin 2\phi & 0 \\
0 & -\sin 2\phi & \cos 2\phi & 0 \\
0 & 0 & 0 & 1
\end{pmatrix}.
\end{equation}

The scattered intensity is given by the first Stokes component, $S_0'(\theta, \phi)$, and directly determines the angular distribution of scattered photons: the probability for scattering into angle $(\theta,\phi)$ is  $p(\theta,\phi)\propto S_0'(\theta, \phi)/S_0$. 

To properly sample the angles, consistently with this distribution, we use the rejection method: We first draw trial values uniformly with 
\(\cos\theta \in [-1,1]\) and \(\phi \in [0,2\pi)\). 
For each trial, we evaluate the probability \(p(\theta,\phi)\), 
and accept the trial if a uniform random number 
\(r \in [0,1]\) is smaller than \(p(\theta,\phi)\), 
otherwise, we generate a new trial.
%\begin{enumerate}
%    \item Propose trial values by drawing uniformly
%    \[
%     \cos\theta \in[-1,1], \qquad \phi \in [0,2\pi).
%    \]
%    \item Compute the  probability  $p(\theta,\phi)$  for the drawn angles
%    \item Draw a uniform random number $r \in [0,1]$.  
%    If $r < p(\theta,\phi)$, accept the direction $(\theta,\phi)$;  
%    otherwise, generate a new trial.
%\end{enumerate}

The comparison between the experimental data of the 3D and quasi-1D clouds with the simulations is addressed through two models: polarization included only in the first scattering event, while all the multiple scattering events are treated as fully depolarized (depolarized model), and polarization accounted for at every scattering event (polarization-tracking model). We consider incident propagation vector $\mathbf{k}_\text{in}=[0,0,1]$, and, as in the experiment, the right-hand circular polarization $\mathbf{S}_\text{in}=[1,0,0,1]$. 
For circularly polarized incident probe light, the scattering probability at the first event is determined by the normalized scattered intensity,  
\begin{equation}
p_1(\theta) = \frac{S_0'(\theta)}{S_0'^{\max}}
= \frac{1+\cos^2\theta}{2}.
\end{equation}  
In the depolarized model, the angles upon first scattering event are sampled with the rejection method according to this probability distribution. For all subsequent events, scattering angles are drawn uniformly: $\cos\theta \in [-1,1], \; \phi \in [0,2\pi)$. To account for experimental polarization filtering at the detection, where the detection optics is aligned to transmit the polarization of the probe, statistical filtering is applied such that each multiply scattered photon is kept or discarded with $50\%$ probability. This ensures that only half of the multiply scattered photons are detected, as they are considered to be fully depolarized in this model.
 
 For the polarization-tracking model, once a trial set of angles $(\theta,\phi)$ is accepted via the rejection method and scattering probability $p(\theta,\phi)$, the photon’s polarization state is updated according to Eq.~\eqref{eq:Stokes_update}. At the same time, the local polarization reference frame, as well as propagation direction $\mathbf{k}$ are updated via Euler rotations determined by $(\theta,\phi)$. The updated polarization state $\mathbf{S'}$ together with the new propagation vector and local basis, serves as the input for the next scattering event in the random walk. This procedure is repeated at every scattering step, until the photon leaves the sample. To reproduce the experimental detection scheme with polarization filtering, the exiting Stokes vector is first rotated into the laboratory detection basis. A circular polarizer is then applied by acting with the corresponding Mueller matrix. Finally, a statistical filtering method is used: each photon is accepted or rejected with a probability proportional to the transmitted intensity $S_0^{\text{out}}$ after the polarizer, so that only photons transmitted by the circular analyzer contribute to the simulated signal, as in the experiment. 

 Fig.~\ref{Fig:Angular_dist} presents the results of RW simulations, showing the angular distribution of scattered photons for two optical thicknesses, comparing quasi-1D and 3D cloud configurations. For each case, we show the contributions from singly scattered photons, multiply scattered photons, and the ratio of multiple to the total scattered light. The simulations also include the two polarization models: depolarized and polarization-tracking. Several key features can be  observed. First, for a given optical thickness, the quasi-1D cloud exhibits less multiply scattered photons than the 3D cloud (Fig.~\ref{Fig:Angular_dist}(b--c, e--f)). This suppression of multiple scattering in elongated geometry reflects the fact that photons scattered in their first scattering event away from the near-forward direction are more likely to leave the elongated cloud without undergoing further scattering events (Fig.~\ref{Fig:Angular_dist}(a,d)). Second, in the polarization-tracking model, the near-forward direction shows a larger contribution of multiple scattering compared to the depolarized model (Fig.~\ref{Fig:Angular_dist}(b--c, e--f)), highlighting the role of polarization in the detection setup. Finally, for the 3D cloud at higher optical thickness, the results from the two models become more similar, as the larger number of scattering events enhances depolarization and drives the polarization-tracking model closer to the depolarized model limit (Fig.~\ref{Fig:Angular_dist}(f)).

As it is outlined in the main text, the experimental data for the 3D cloud match the predictions of the depolarized model, whereas in the quasi-1D geometry the results align more closely with polarization-tracking model of RW simulations. One reason is that photons undergo a larger number of scattering events in the 3D case, which naturally enhances depolarization. Two extra experimental factors contribute to the stronger depolarization in 3D configuration: (i) the $|F=3\rangle \to |F=4^\prime\rangle$ transition of $^{85}$Rb involves many Zeeman sublevels with different Clebsch–Gordan coefficients, unlike the closed cycling transition used in RW simulations. This allows reabsorption and re-emission on different sublevels, thereby increasing polarization mixing of multiply scattered photons. (ii)  Partial optical pumping, as the probe beam, with the waist size much smaller than the size of the 3D cloud, optically pumps only the central region of the cloud into the stretched state, leaving atoms in the outer regions distributed across different Zeeman sublevels. Together, these factors accelerate depolarization, effectively rendering multiple scattering fully depolarized in the 3D geometry.

\end{document}